\documentclass[reqno,a4paper,final]{amsart}
\usepackage{amssymb,amstext,amsthm,eucal}
\usepackage{bbold}
\usepackage{array}
\usepackage[utf8]{inputenc}
\usepackage{hyperref}
\hypersetup{%
  pdftitle = {Supersymmetry and homogeneity of M-theory backgrounds},
  pdfkeywords = {supergravity, Killing spinors, Killing vectors,
    isometries, supersymmetry, superalgebra, Lie superalgebra, Killing
    superalgebra},
  pdfauthor = {José Figueroa-O'Farrill, Patrick Meessen, Simon Philip},
  pdfcreator = {\LaTeX\ with package \flqq hyperref\frqq} }
\newcommand{\half}{\tfrac{1}{2}}

\newcommand{\fg}{\mathfrak{g}}
\newcommand{\fh}{\mathfrak{h}}
\newcommand{\fk}{\mathfrak{k}}

\newcommand{\fM}{\mathfrak{M}}

\newcommand{\fosp}{\mathfrak{osp}}
\newcommand{\fsl}{\mathfrak{sl}}
\newcommand{\fso}{\mathfrak{so}}
\newcommand{\fsp}{\mathfrak{sp}}

\newcommand{\fsu}{\mathfrak{su}}
\newcommand{\fu}{\mathfrak{u}}

\newcommand{\SO}{\mathrm{SO}}

\newcommand{\Cl}{\mathrm{C}\ell}
\newcommand{\Spin}{\mathrm{Spin}}
\ifx\Sp\UnDeFiNeD
\newcommand{\Sp}{\mathrm{Sp}}
\else
\renewcommand{\Sp}{\mathrm{Sp}}
\fi

\newcommand{\SU}{\mathrm{SU}}
\newcommand{\U}{\mathrm{U}}

\newcommand{\RR}{\mathbb{R}}

\newcommand{\ZZ}{\mathbb{Z}}

\newcommand{\eD}{\mathcal{D}}
\newcommand{\eE}{\mathcal{E}}

\newcommand{\eL}{\mathcal{L}}

\newcommand{\eS}{\mathcal{S}}

\DeclareMathOperator{\AdS}{AdS}

\DeclareMathOperator{\im}{im}
\DeclareMathOperator{\rank}{rank}
\DeclareMathOperator{\End}{End}

\newcommand{\bx}{\boldsymbol{x}}

\newcommand{\1}{\mathbb{1}}

\newcommand{\repre}[1]{\underline{\mathbf{#1}}}

\newcommand{\MUNCH}[1]{\relax}

\allowdisplaybreaks[1]
\begin{document}
\title[Supersymmetry and homogeneity]{Supersymmetry and homogeneity
  of M-theory backgrounds}
\author[Figueroa-O'Farrill]{José Figueroa-O'Farrill}
\author[Meessen]{Patrick Meessen}
\author[Philip]{Simon Philip}
\address[JMF,SP]{School of Mathematics, University of Edinburgh}
\email{J.M.Figueroa@ed.ac.uk, S.A.R.Philip@sms.ed.ac.uk}
\address[PM]{Theory Group, CERN, Geneva, Switzerland}
\email{Patrick.Meessen@cern.ch}
\thanks{EMPG-04-09, CERN-PH-TH/2004-145}
\begin{abstract}
  We describe the construction of a Lie superalgebra associated to an
  arbitrary supersymmetric M-theory background, and discuss some
  examples.  We prove that for backgrounds with more than 24
  supercharges, the bosonic subalgebra acts locally transitively.  In
  particular, we prove that backgrounds with more than 24
  supersymmetries are necessarily (locally) homogeneous.
\end{abstract}
\maketitle
\tableofcontents

\section{Introduction}
\label{sec:intro}

The amount of preserved supersymmetry is an important invariant of a
supergravity background; one which has played a pivotal role in the
investigations on duality in string theory.  This invariant, usually
specified as a fraction $\nu$ of the supersymmetry of the theory,
admits two complementary refinements: the holonomy representation of
the superconnection defined by the variation of the gravitino on the
one hand, and the supersymmetry superalgebra on the other.  One can
recover $\nu$ from either of these two: from the dimension of the
invariant subspace in the holonomy representation, or from the
dimension of the odd subspace in the superalgebra.  The holonomy
representation and the supersymmetry superalgebra are not unrelated
\cite{DuffLiuHol,HullHolonomy}; although precisely what this relation
is remains to be elucidated.

Concentrating on the supersymmetry superalgebra for a moment, the
seemingly trivial fact that supersymmetries give rise to symmetries
suggests that the more supersymmetric a background, the more
`symmetric' it ought to be.  Indeed, the maximally supersymmetric
backgrounds are all symmetric spaces \cite{FOPMax}, hence in
particular they are homogeneous.  Is is therefore natural to ask how
much supersymmetry must a background preserve for it to be
automatically homogeneous.  In other words, is there a critical
fraction $\nu_c$, such that if a background preserves a fraction $\nu
> \nu_c$ of the supersymmetry then it is guaranteed to be homogeneous?

Let us concentrate for definiteness on M-theory backgrounds; that is,
bosonic solutions of the equations of motion of eleven-dimensional
supergravity \cite{Nahm,CJS}.  Based on known examples and some
indirect arguments that we will review presently, a natural conjecture
might be that $\nu_c = \half$.  Indeed, there are plenty of
$\nu=\half$ backgrounds which are \emph{not} homogeneous, for example,
the elementary $\half$-BPS backgrounds: the generic M-wave
\cite{MWave}, the M-branes \cite{DS2brane,Guven} and the Kaluza--Klein
monopole \cite{SMKK,GPMKK,HKMKK} are not homogeneous; whence $\nu_c
\geq \half$.  On the other hand, all known examples of backgrounds
with $\nu>\half$, hereafter denoted 16+, are homogeneous.  These
examples include the maximally supersymmetric solutions: flat space,
the Freund--Rubin backgrounds $\AdS_4 \times S^7$ and $S^4 \times
\AdS_7$ \cite{FreundRubin}, and the Kowalski-Glikman wave
\cite{KG,FOPFlux}; discrete cyclic quotients \cite{FigSimAdS} of
$\AdS_4 \times S^7$ with $\nu = \frac34$ and $\nu = \frac9{16}$; the
Gödel backgrounds \cite{GGHPR,HarTakaGoedel}; and a number of plane
waves, both symmetric
\cite{CLPpp,Michelson,ChrisJerome,CLPppM,Michelson26} and
time-dependent \cite{BMO}.  The proof of the homogeneity of the
discrete cyclic quotients is presented in
Appendix~\ref{app:homogquots} for the first time, whereas
Appendix~\ref{app:newwave} contains a novel non-symmetric plane wave
solution with $22$ supercharges.

An indirect argument supporting the $\nu_c = \half$ ``conjecture''
would be that 16+ plane waves are known to be homogeneous \cite{BMO}.
However as illustrated in \cite{Patricot}, homogeneity is \emph{not} a
hereditary property \cite{Geroch} of the plane-wave limit
\cite{PenrosePlaneWave,GuevenPlaneWave}.  Therefore one cannot argue
that their homogeneity can be explained \emph{a posteriori} by the
hereditary property of supersymmetry \cite{Limits}, just like the
plane-wave limit explains \cite{ShortLimits} the existence of the
maximally supersymmetric plane waves\cite{KG,NewIIB,Meessen}.

In this paper we will prove that every 24+ M-theory background is
(locally) homogeneous.

This paper is organised as follows.  After briefly reviewing the
definition of a supersymmetric M-theory background in
Section~\ref{sec:susybgs} and Kostant's approach to Killing vectors in
\ref{sec:isometries}, we define the Killing superalgebra of a
supersymmetric M-theory background and prove that it is always a Lie
superalgebra.  Special cases of this construction have appeared in
\cite{AFHS,GMT1,GMT2,PKT,JMFKilling,FOPFlux,NewIIB,ALOKilling}, but in
Section~\ref{sec:superalgebra} we treat the general case.  In
Section~\ref{sec:examples} we compute the Killing superalgebra of some
standard backgrounds.  In Section~\ref{sec:susyhom} we define
different notions of homogeneity for supergravity backgrounds and
prove that 24+ backgrounds are locally homogeneous. In Section~\ref{sec:conc} we offer some conclusions.
The paper ends with several appendices.  Appendix~\ref{app:clifford}
contains our conventions for the Clifford algebra needed for the
calculations in this paper.  Appendix~\ref{app:homogquots} discusses
the homogeneity and the Killing superalgebras of some 16+ discrete
quotients, whereas Appendix~\ref{app:newwave} presents a new
time-dependent 16+ homogeneous plane wave.

\section{Supersymmetric M-theory backgrounds}
\label{sec:susybgs}

Let $(M,g,F)$ be a classical M-theory background, where $(M,g)$ is a
connected eleven-dimensional lorentzian spin manifold and $F$ is a
closed four-form, subject to the well-known field equations whose
explicit form are of no consequence in what follows.

Let $\eS$ denote the bundle of spinors of this background.  (Our
Clifford conventions are explained in Appendix~\ref{app:clifford}.)
The bundle $\eS$ is a bundle of Clifford modules, modelled locally on
an irreducible $\Cl(1,10)$-module.  There are two such modules up to
isomorphism: they are both real and 32-dimensional and are
distinguished by the action of the centre of $\Cl(1,10)$, which is
generated by the volume form.  Our formulae are valid for the Clifford
module on which the action of the centre of $\Cl(1,10)$ is nontrivial.
There is an equivalent version of the theory for the other choice of
Clifford module, in which the supersymmetry transformations will
differ by some signs.

The variation of the gravitino, after setting the gravitino to zero,
defines a connection $\eD$ on $\eS$, given by
\begin{equation}
  \label{eq:connection}
  \eD_X = \nabla_X  + \tfrac16 \iota_X F + \tfrac1{12} X^\flat \wedge
  F~,
\end{equation}
where $X^\flat$ is the one-form dual to $X$ and differential forms act
on spinors via the Clifford action as reviewed in
Appendix~\ref{app:clifford}.  The difference in sign from, say,
\cite{Bonn} is due to our using a mostly minus metric, which changes
the sign of the musical isomorphism $\flat$.

Nonzero sections of $\eS$ which are parallel relative to $\eD$ are
called \textbf{Killing spinors}, and an M-theory background $(M,g,F)$
admitting Killing spinors is said to be \textbf{supersymmetric}.
Since $M$ is connected, a Killing spinor is uniquely determined by its
value at a point: its value at any other point is obtained by parallel
translating with respect to the connection $\eD$.  Since Killing
spinors are parallel with respect to $\eD$, this does not depend on
the path.  Killing spinors therefore define a real sub-bundle $W
\subset \eS$, where for $p\in M$, $W_p \subset \eS_p$ is the subspace
spanned by the values of all Killing spinors at $p$.  The rank of $W$
is equal to $32 \nu$, where $\nu$ is the \textbf{fraction of
  supersymmetry} preserved by the background.

\section{Killing vectors}
\label{sec:isometries}

Whereas Killing spinors are uniquely determined by their value at a
point, to specify a Killing vector one requires its value at a point
\emph{and} that of its first derivative.  Indeed, as explained for
example in \cite{KostantHol,Geroch}, Killing vectors are in one-to-one
correspondence with parallel sections of the bundle
\begin{equation*}
  \eE = TM \oplus \fso(TM)~,
\end{equation*}
where $\fso(TM)$ is the bundle of skew-symmetric endomorphisms
(relative to $g$) of the tangent bundle, and where the connection is
the one defining the so-called Killing transport
\cite{KostantHol,Geroch}.  Let us review this now.

Let $(M,g)$ be a connected pseudo-riemannian manifold and $\xi$ a
vector field.  Let $A_\xi: TM \to TM$ be defined by $A_\xi X = -
\nabla_X \xi$.  Then $\xi$ is a Killing vector if and only if $A_\xi$
is skew-symmetric relative to the metric, denoted here
$\left<-,-\right>$,
\begin{equation*}
  \left<A_\xi X, Y\right> = - \left<A_\xi Y, X\right>~.
\end{equation*}

Killing's identity says that
\begin{equation}
  \label{eq:killing}
  \nabla_X A_\xi = R(X, \xi)~,
\end{equation}
where $R(X,Y) : TM \to TM$ is defined by
\begin{equation*}
  R(X,Y) Z = \nabla_{[X,Y]}Z - \nabla_X\nabla_Y Z + \nabla_Y\nabla_X
  Z~.
\end{equation*}

\begin{proof}
  Notice that
  \begin{align*}
    (\nabla_X A_\xi)Y &= \nabla_X A_\xi Y - A_\xi \nabla_X Y\\
    &= - \nabla_X \nabla_Y \xi + \nabla_{\nabla_X Y} \xi~,
  \end{align*}
  whence
  \begin{align*}
    (\nabla_X A_\xi)Y - (\nabla_Y A_\xi)X &= - \nabla_X \nabla_Y
    \xi + \nabla_{\nabla_X Y} \xi
    +  \nabla_Y \nabla_X \xi - \nabla_{\nabla_Y X} \xi\\
    &= - \nabla_X \nabla_Y \xi + \nabla_Y \nabla_X \xi +
    \nabla_{[X,Y]}\xi\\
    & = R(X,Y) \xi\\
    &= R(X, \xi)Y - R(Y, \xi)X~,
  \end{align*}
  where we have used the algebraic Bianchi identity
  \begin{equation}
    \label{eq:bianchi}
    R(X,Y)Z + R(Y,Z)X + R(Z,X)Y = 0~.
  \end{equation}
  This means that
  \begin{equation*}
    (\nabla_X A_\xi)(Y) - R(X, \xi)Y
  \end{equation*}
  is symmetric in $X \leftrightarrow Y$.  On the other hand,
  \begin{equation*}
    \left<(\nabla_X A_\xi)Y - R(X, \xi)Y, Z\right> = 
    -  \left<(\nabla_X A_\xi)Z - R(X,\xi)Z, Y\right>~,
  \end{equation*}
  whence $\left<(\nabla_X A_\xi)Y - R(X,\xi)Y, Z\right> = 0$.
\end{proof}

This means that a Killing vector $\xi$ is uniquely characterised by
the data
\begin{equation*}
  (\xi_p, -\nabla\xi_p) \in T_pM \oplus \fso(T_pM)~,
\end{equation*}
at any point $p\in M$.  Indeed, Killing vectors are in one-to-one
correspondence with parallel sections of the bundle $\eE = TM \oplus
\fso(TM)$ under the connection
\begin{equation*}
  D_X
  \begin{pmatrix}
    \xi \\ A
  \end{pmatrix}
  =
  \begin{pmatrix}
    \nabla_X\xi + A(X)\\ \nabla_XA - R(X,\xi)
  \end{pmatrix}~.
\end{equation*}

Let $\fk$ denote the space of parallel sections of $\eE$. The Lie
bracket of Killing vectors induces a Lie algebra structure on $\fk$ as
follows.  Let $(\xi,A)$ and $(\eta, B)$ be parallel sections.  Their
Lie bracket is given by
\begin{equation}
  \label{eq:g0bracket}
  [(\xi,A),(\eta,B)] = (A \eta - B \xi, [A,B] + R(\xi,\eta))~.
\end{equation}
\begin{proof}
  By definition,
  \begin{equation*}
    [(\xi,A),(\eta,B)] = ([\xi,\eta], -\nabla[\xi,\eta])~.
  \end{equation*}
  Now, the torsionless condition of $\nabla$ means that
  \begin{equation*}
    [\xi,\eta] = \nabla_\xi \eta - \nabla_\eta \xi = A \eta - B \xi~,
  \end{equation*}
  using that $A = - \nabla\xi$ and $B = - \nabla\eta$.  Similarly,
  \begin{align*}
    -\nabla_X [\xi,\eta] &= -\nabla_X (A \eta - B \xi)\\
    &= -(\nabla_X A) \eta - A \nabla_X \eta + (\nabla_X B)\xi + B
    (\nabla_X \xi)\\
    &= - R(X,\xi) \eta + A B X + R(X,\eta)\xi - B A X\\
    &= [A,B]X + R(\xi,\eta)X~,
  \end{align*}
  where we have used Killing's identity \eqref{eq:killing} and the
  algebraic Bianchi identity \eqref{eq:bianchi}.
\end{proof}
Now the bundle $\eE$ is naturally a bundle of Lie algebras with Lie
bracket
\begin{equation*}
  [(\xi,A),(\eta,B)]_{\eE} = (A \eta - B \xi, [A,B])~.
\end{equation*}
Therefore we see that the curvature $R(\xi,\eta)$ measures the failure
of this natural Lie bracket to agree with the Lie bracket in $\fk$.
Indeed, the bracket on $\fk$ extends to arbitrary sections of $\eE$,
but it will fail to satisfy the Jacobi identity precisely due to the
curvature term.

If $(M,g,F)$ is a supergravity background, then the $F$-preserving
elements of $\fk$ define a Lie subalgebra which, anticipating our next
topic, will be denoted $\fg_0$.

\section{The Killing superalgebra}
\label{sec:superalgebra}

The Killing spinors and the $F$-preserving Killing vectors of a
supergravity background $(M,g,F)$ define a Lie superalgebra, which we
call the \textbf{Killing superalgebra} of the background.

We shall denote the Killing superalgebra by $\fg = \fg_0 \oplus
\fg_1$, where the even subalgebra $\fg_0$ is the Lie algebra of
$F$-preserving Killing vectors and the odd subspace $\fg_1$ consists
of (the ``oddification'' of) the Killing spinors.  The grading implies
that we must distinguish three types of brackets.

First of all we have the bracket $[-,-]: \fg_0 \otimes \fg_0 \to
\fg_0$, corresponding to the Lie bracket of Killing vectors defined in
\eqref{eq:g0bracket}.  It clearly satisfies the Jacobi identity,
whence $\fg_0$ is a Lie algebra.

The bracket $[-,-]: \fg_0 \otimes \fg_1 \to \fg_1$ corresponds to the
action of the Killing vectors on the Killing spinors via the spinorial
Lie derivative \cite{Kosmann}.  Let $\rho: \fso(TM) \to \End \eS$
denote the spinor representation.  Then if $(\xi,A_\xi) \in \fk$, and
$\varepsilon \in \fg_1$, we define
\begin{equation}
  \label{eq:Lieder}
  [(\xi,A_\xi), \varepsilon] = \nabla_\xi \varepsilon + \rho(A_\xi)
  \varepsilon~,
\end{equation}
where the right-hand side defines the \textbf{spinorial Lie
  derivative} $\eL_\xi$.  If $(\xi,A_\xi) \in \fg_0$, then the
right-hand side will again be in $\fg_1$ since for all vector fields
$X$, one has
\begin{equation*}
  [\eL_\xi, \eD_X] = \eD_{[\xi,X]}~.
\end{equation*}
The spinorial Lie derivative satisfies
\begin{equation}
  \label{eq:001jacobi}
  \eL_X \eL_Y \varepsilon -  \eL_Y \eL_X \varepsilon = \eL_{[X,Y]}
  \varepsilon~.
\end{equation}
\begin{proof}
  Applying \eqref{eq:Lieder} and dropping $\rho$ from the notation,
  we find
  \begin{align*}
    [\eL_X, \eL_Y]\varepsilon &= \eL_X (\nabla_Y \varepsilon +
    A_Y\varepsilon) - \eL_Y (\nabla_X \varepsilon +
    A_X\varepsilon)\\
    &= \nabla_X \nabla_Y \varepsilon + A_X \nabla_Y \varepsilon +
    \nabla_X( A_Y \varepsilon) + A_X A_Y \varepsilon - (X
    \leftrightarrow Y)\\
    &= \nabla_X \nabla_Y \varepsilon - \nabla_Y \nabla_X \varepsilon
    + [A_X, A_Y] \varepsilon + (\nabla_X A_Y) \varepsilon -
    (\nabla_Y A_X) \varepsilon~.
  \end{align*}
  We now use that
  \begin{equation*}
    [\nabla_X,\nabla_Y] \varepsilon = \nabla_{[X,Y]} \varepsilon -
    R(X,Y) \varepsilon
  \end{equation*}
  and Killing's identity \eqref{eq:killing} repeatedly to arrive at
  \begin{align*}
    [\eL_X, \eL_Y]\varepsilon &= \nabla_{[X,Y]} \varepsilon +
    [A_X,A_Y] \varepsilon + R(X,Y)\varepsilon\\
    &= \nabla_{[X,Y]} \varepsilon + A_{[X,Y]} \varepsilon\\
    &= \eL_{[X,Y]} \varepsilon~.
  \end{align*}
\end{proof}
Equation \eqref{eq:001jacobi} is equivalent to the
$[\fg_0,\fg_0,\fg_1]$-Jacobi identity.

The bracket $[-,-]: \fg_1 \otimes \fg_1 \to \fg_0$ is induced from the
tensor-square of the corresponding Killing spinors.  Indeed, we have
a map
\begin{equation}
  \label{eq:square}
  \xi: \eS \otimes \eS \to TM
\end{equation}
which takes two spinors $\varepsilon_1$ and $\varepsilon_2$ and
produces a vector field $\xi(\varepsilon_1,\varepsilon_2)$ defined
as the unique vector field such that for all other vector fields $Y$,
\begin{equation}
  \label{eq:Xi}
  \left<\xi(\varepsilon_1,\varepsilon_2),Y\right> = (\varepsilon_1,
  Y^\flat \cdot \varepsilon_2)~.
\end{equation}
The map \eqref{eq:square} is defined on all spinors, but its
restriction to Killing spinors has a crucial property: namely, that
if $\varepsilon_1$ and $\varepsilon_2$ are Killing spinors, then
$X = \xi(\varepsilon_1, \varepsilon_2)$ is a Killing vector, so that
$\eL_X g = 0$, which in addition \cite{GauPak} preserves $F$. 
\begin{proof}
  Let $\varepsilon_i$, $i=1,2$, be Killing spinors.  Then for all
  vectors $X,Y$, we have
  \begin{align*}
    \left<\nabla_X \xi(\varepsilon_1, \varepsilon_2), Y\right> &= X
    \left<\xi(\varepsilon_1, \varepsilon_2), Y\right> -
    \left<\xi(\varepsilon_1, \varepsilon_2), \nabla_X Y\right>\\
    &= X (\varepsilon_1, Y^\flat\cdot \varepsilon_2) - (\varepsilon_1,
    \nabla_X Y^\flat\cdot \varepsilon_2)\\
    &= (\nabla_X \varepsilon_1, Y^\flat\cdot \varepsilon_2)
    + (\varepsilon_1, Y^\flat \cdot \nabla_X \varepsilon_2)~.
  \end{align*}
  Using that $\eD\varepsilon_i=0$, we can rewrite this as
  \begin{equation*}
    \left<\nabla_X \xi(\varepsilon_1, \varepsilon_2), Y\right> =
    (\varepsilon_1, \Omega_X^* \cdot Y^\flat\cdot \varepsilon_2)
    + (\varepsilon_1, Y^\flat\cdot \Omega_X \cdot \varepsilon_2)~,
  \end{equation*}
  where
  \begin{equation*}
    \Omega_X = - \tfrac1{12} X^\flat \wedge F -  \tfrac16 \iota_X F~,
  \end{equation*}
  and $\Omega_X^* = \tfrac1{12} X^\flat \wedge F - \tfrac16 \iota_X F$
  is its symplectic adjoint as defined in \eqref{eq:sympadj}.  Using
  equations \eqref{eq:Xomega} and \eqref{eq:omegaX} in
  Appendix~\ref{app:clifford}, we arrive at
  \begin{equation}
    \label{eq:NablaXi}
    \left<\nabla_X \xi(\varepsilon_1, \varepsilon_2), Y\right> =
    - \tfrac13 (\varepsilon_1, \iota_X \iota_Y F \cdot \varepsilon_2)
    + \tfrac16 (\varepsilon_1, X^\flat \wedge Y^\flat \wedge F \cdot
    \varepsilon_2)~,
  \end{equation}
  which is manifestly skew-symmetric in $X$ and $Y$, showing that
  $\xi(\varepsilon_1, \varepsilon_2)$ is a Killing vector.
  
  Now define a 2-form $B$ by
  \begin{equation}
    \label{eq:spinor2form}
    B(X,Y) = (\varepsilon_1, X^\flat \wedge Y^\flat \cdot
    \varepsilon_2)~,
  \end{equation}
  and let us compute its covariant derivative.  By definition,
  \begin{align*}
    (\nabla_Z B)(X,Y) &= (\nabla_Z \varepsilon_1, X^\flat \wedge
    Y^\flat \cdot \varepsilon_2) + (\varepsilon_1, X^\flat \wedge
    Y^\flat \cdot \nabla_Z \varepsilon_2)\\
    &= (\Omega_Z \varepsilon_1, X^\flat \wedge
    Y^\flat \cdot \varepsilon_2) + (\varepsilon_1, X^\flat \wedge
    Y^\flat \cdot \Omega_Z \varepsilon_2)\\
    &= (\varepsilon_1, \Omega_Z^* \cdot (X^\flat \wedge
    Y^\flat) \cdot \varepsilon_2) + (\varepsilon_1, (X^\flat \wedge
    Y^\flat) \cdot \Omega_Z \varepsilon_2)~.
  \end{align*}
  Using equations \eqref{eq:omegaXY} and \eqref{eq:XYomega} in
  Appendix~\ref{app:clifford}, we arrive at
  \begin{multline*}
    (\nabla_Z B)(X,Y) = 
    \tfrac16 g(Y,Z) (\varepsilon_1, X^\flat \wedge F \cdot
    \varepsilon_2) - \tfrac16 g(X,Z) (\varepsilon_1, Y^\flat \wedge F
    \cdot \varepsilon_2)\\
    + \tfrac16 (\varepsilon_1, Y^\flat \wedge
    Z^\flat \wedge \iota_X F \cdot \varepsilon_2) +
    \tfrac16 (\varepsilon_1, Z^\flat \wedge
    X^\flat \wedge \iota_Y F \cdot \varepsilon_2)\\
    - \tfrac13 (\varepsilon_1, X^\flat \wedge
    Y^\flat \wedge \iota_Z F \cdot \varepsilon_2)
    - \tfrac13 (\varepsilon_1, \iota_X\iota_Y\iota_Z F \cdot
    \varepsilon_2)~.
  \end{multline*}
  We now alternate this equation to obtain $dB$:
  \begin{align*}
    dB(X,Y,Z) &= (\nabla_X B)(Y,Z) + (\nabla_Y B)(Z,X) + (\nabla_Z
    B)(X,Y)\\
    &= - (\varepsilon_1, \iota_X \iota_Y \iota_Z F \cdot
    \varepsilon_2)~.
  \end{align*}
  Noticing that
  \begin{equation*}
    (\varepsilon_1, \iota_X \iota_Y \iota_Z F \cdot
    \varepsilon_2) = F(\xi(\varepsilon_1, \varepsilon_2), X, Y, Z)~,
  \end{equation*}
  we have that
  \begin{equation*}
    \iota_{\xi(\varepsilon_1, \varepsilon_2)} F = - dB~.
  \end{equation*}
  Since $F$ is closed, this implies that the vector field
  $\xi(\varepsilon_1, \varepsilon_2)$ leaves $F$ invariant.
\end{proof}
It is convenient to extend the map $\xi$ to a map
\begin{equation}
  \label{eq:varphi}
  \varphi : \eS \otimes \eS \to \eE
\end{equation}
This maps restricts to a map sending parallel sections (with respect
to $\eD$) of $\eS \otimes \eS$ to parallel sections (with respect to
$D$) of $\eE$, which we will also denote $\varphi$.  The explicit form
of this map is given by
\begin{equation*}
  \varphi(\varepsilon_1, \varepsilon_2) =
  (\xi(\varepsilon_1,\varepsilon_2), - \nabla
  \xi(\varepsilon_1,\varepsilon_2))~,
\end{equation*}
where $\xi(\varepsilon_1,\varepsilon_2)$ and $\nabla
\xi(\varepsilon_1,\varepsilon_2)$ are given by equations \eqref{eq:Xi}
and \eqref{eq:NablaXi}, respectively.  The fundamental property of the
map $\varphi$ is its equivariance under the action of $\fg_0$.  In
other words,
\begin{equation}
  \label{eq:uivariance}
  [(X,A_X) , \varphi(\varepsilon_1,\varepsilon_2)] = 
  \varphi(\eL_X \varepsilon_1,\varepsilon_2) +
  \varphi(\varepsilon_1, \eL_X \varepsilon_2)~.
\end{equation}
Equivalently, for all vector fields $Y$ (not necessarily Killing),
\begin{equation*}
  \left<\eL_X \xi(\varepsilon_1,\varepsilon_2), Y\right> = (\eL_X
  \varepsilon_1, Y^\flat \cdot \varepsilon_2) + 
  (\varepsilon_1, Y^\flat \cdot \eL_X \varepsilon_2)~.
\end{equation*}
\begin{proof}
  Computing the left-hand side, we find
  \begin{align*}
    \left<\eL_X \xi(\varepsilon_1,\varepsilon_2), Y\right> &= 
    \left<\nabla_X \xi(\varepsilon_1,\varepsilon_2) -
    \nabla_{\xi(\varepsilon_1,\varepsilon_2)} X, Y\right>\\
    &= (\nabla_X \varepsilon_1, Y^\flat\cdot \varepsilon_2) +
    (\varepsilon_1, Y^\flat\cdot \nabla_X \varepsilon_2) +
    (\varepsilon_1, \nabla_Y X^\flat\cdot \varepsilon_2)~.
  \end{align*}
  Computing the right-hand side, we obtain
  \begin{align*}
    (\eL_X \varepsilon_1, Y^\flat\cdot \varepsilon_2) +
    (\varepsilon_1, Y^\flat\cdot \eL_X \varepsilon_2) &=
    (\nabla_X \varepsilon_1, Y^\flat\cdot \varepsilon_2) +
    (A_X \varepsilon_1, Y^\flat\cdot \varepsilon_2)\\
    & \qquad + (\varepsilon_1, Y^\flat\cdot \nabla_X \varepsilon_2) +
    (\varepsilon_1, Y^\flat\cdot A_X \cdot \varepsilon_2)~.
  \end{align*}
  The difference is therefore
  \begin{align*}
    (\varepsilon_1, \nabla_Y X^\flat\cdot \varepsilon_2) +
    (\varepsilon_1, A_X \cdot Y^\flat\cdot \varepsilon_2) -
    (\varepsilon_1, Y^\flat\cdot A_X \cdot \varepsilon_2)~,
  \end{align*}
  which is easily seen to vanish as a consequence of the identity
  \begin{equation*}
    [A_X, Y] = A_X(Y) = - \nabla_Y X ~.
  \end{equation*}
\end{proof}
Equation \eqref{eq:uivariance} is precisely the
$[\fg_0,\fg_1,\fg_1]$-Jacobi identity.  It also implies that
$[\fg_1,\fg_1]\subset \fg_0$ is an ideal, which is a general fact of
superalgebras.  In other words, $\fg_1$ generates an ideal
$[\fg_1,\fg_1] \oplus \fg_1 \subset \fg$.

Finally we consider the $[\fg_1,\fg_1,\fg_1]$-Jacobi identity.  This
is equivalent to the vanishing of a $\fg_0$-equivariant symmetric
trilinear map $J: S^3\fg_1 \to \fg_1$, defined by
\begin{equation}
  \label{eq:jacobator}
  J(\varepsilon_1,\varepsilon_2,\varepsilon_3) :=
  \eL_{\xi(\varepsilon_1,\varepsilon_2)} \varepsilon_3 + 
  \eL_{\xi(\varepsilon_2,\varepsilon_3)} \varepsilon_1 +
  \eL_{\xi(\varepsilon_3,\varepsilon_1)} \varepsilon_2~.
\end{equation}
The vanishing of $J$ is equivalent to
\begin{equation}
  \label{eq:FFFjac}
  \eL_{\xi(\varepsilon,\varepsilon)} \varepsilon = 0~,
\end{equation}
for all Killing spinors $\varepsilon$.
\begin{proof}
  Equation \eqref{eq:FFFjac} is simply
  $J(\varepsilon,\varepsilon,\varepsilon) = 0$, up to an overall
  factor of $3$.  Hence this vanishes when $J$ vanishes.  Conversely
  we can use the standard polarisation tricks; that is, apply
  \eqref{eq:FFFjac} to $\varepsilon= \varepsilon_1 + \varepsilon_2 +
  \varepsilon_3$ to obtain that $2 J(\varepsilon_1, \varepsilon_2,
  \varepsilon_3) = 0$.
\end{proof}
In other words, the Jacobi identity is equivalent to every Killing
spinor being left invariant by the Killing vector obtained by 
squaring it.

Equation \eqref{eq:FFFjac} does not involve any derivatives.
Indeed, it is equivalent to
\begin{equation}
  \label{eq:jacobator2}
  \left( 2 \iota_\xi F + \xi^\flat \wedge F + B \wedge
    \star F + C \wedge F \right) \cdot \varepsilon = 0~,
\end{equation}
where $\xi^\flat$, $B$ and $C$ are the 1-, 2- and 5-forms constructed
out of the Killing spinor $\varepsilon$, respectively:
\begin{gather*}
  \xi^\flat(X) = \left(\varepsilon, X^\flat \cdot \varepsilon\right)\\
  B(X,Y) = \left(\varepsilon, X^\flat \wedge Y^\flat \cdot
    \varepsilon\right)\\
  C(X_1,\dots,X_5) = \left(\varepsilon, X_1^\flat \wedge \cdots \wedge
    X_5^\flat \cdot \varepsilon\right)~.
\end{gather*}
Equation \eqref{eq:FFFjac} is clearly linear in $F$ and cubic in
$\varepsilon$ and furthermore it is equivariant under the action of
$\Spin(1,10)$.  As a consequence, it need only be checked for one
$(F,\varepsilon)$ in each of the (projectivised) $\Spin(1,10)$-orbits
of the relevant representation space.  Rather than working out the
orbit decomposition of this rather large space, we can instead try to
prove that this identity holds for \emph{all} $F$ and for one spinor
$\varepsilon$ in each of the (projectivised) $\Spin(1,10)$ orbits in
the spinor representation.  There are two such orbits, distinguished
by the causal character of the Killing vector associated with
$\varepsilon$.  This can be checked by computer using an explicit real
realisation of $\Cl(1,10)$.  For this it is convenient to unpack
\eqref{eq:jacobator2} further and rewrite it as
\begin{multline}
  \label{eq:explicit}
  F_{abcd} \biggl( \tfrac23 (\varepsilon,\Gamma^a\varepsilon) \Gamma^{bcd}
    + \tfrac1{12} (\varepsilon,\Gamma_e\varepsilon)
    \Gamma^{abcde}\\
    + (\varepsilon,\Gamma^{ab}\varepsilon)
    \Gamma^{cd} + \tfrac1{24}
    (\varepsilon,\Gamma^{abcdmn}\varepsilon) \Gamma_{mn}\biggr) \varepsilon
    = 0~,
\end{multline}
where we have used the Einstein summation convention.  Equation
\eqref{eq:explicit} has been shown to hold for all $F$ and all
$\varepsilon$ using two independent computer calculations: one in
Maple and one in Mathematica.  The relevant code is available upon
request from the authors.

\section{Some examples}
\label{sec:examples}

In this section we will discuss several examples of Killing
superalgebras for some M-theory backgrounds.

\subsection{Purely gravitational backgrounds}
\label{sec:grav}

We start with those backgrounds where $F=0$.  In this case the Killing
spinors are parallel relative to the Levi-Cività connection.  This
means that so are the vectors in $[\fg_1,\fg_1]$.  In particular,
their action on $\fg_1$ is trivial.  This means that $[\fg_1,\fg_1]$
is abelian and, for the purely gravitational backgrounds, they consist
of translations.

Examples of such backgrounds are flat space, the M-wave \cite{MWave},
the Kaluza--Klein monopole \cite{SMKK,GPMKK,HKMKK} as well as their
generalisations \cite{JMWaves}.  For flat space, $[\fg_1,\fg_1]$
coincides with the translation ideal.  For the M-wave, we obtain a
one-dimensional ideal spanned by the parallel null vector $v$ in the
pp-wave.  Indeed, let $u$ be a complementary null vector such that $u
\cdot v + v \cdot u = \1$ in the Clifford algebra.  Such a vector
always exists locally.  The Killing spinors $\varepsilon$ satisfy the
condition $v \cdot \varepsilon = 0$, which means that $\varepsilon = v
\cdot u \cdot \varepsilon$.  Now let $\varepsilon_1,\varepsilon_2$ be
Killing vectors.  If $X \perp v$ then
\begin{align*}
  \left<\xi(\varepsilon_1, \varepsilon_2), X \right> & =
  (\varepsilon_1, X \cdot \varepsilon_2)\\
  &=   (\varepsilon_1, X \cdot v \cdot u \cdot \varepsilon_2)\\
  &= - (\varepsilon_1, v \cdot X \cdot u \cdot \varepsilon_2)\\
  &= ( v \cdot \varepsilon_1, X \cdot u \cdot \varepsilon_2)\\
  &= 0~.
\end{align*}
Therefore $\xi(\varepsilon_1,\varepsilon_2)$ is perpendicular to every
vector $X$ which is perpendicular to $v$, whence it is collinear with
$v$.  Since $v$ and $\xi(\varepsilon_1,\varepsilon_2)$ are both
parallel, we see that $\xi(\varepsilon_1,\varepsilon_2) = c v$ for
some constant $c$.

For the Kaluza--Klein monopole and its generalisations, we obtain the
translations in the flat factor.  Indeed, the geometry here is
$\RR^{1,10-n} \times X^n$ where $X$ is a riemannian manifold
admitting parallel spinors and having no flat directions; that is, no
parallel vector fields.  The possible holonomy groups of $X$ are
tabulated in \cite{Wang} and are given by $\SU(5)$ for $n=10$, any of
$\Sp(1) \times \Sp(1) \subset \Sp(2) \subset \SU(4) \subset \Spin(7)$
for $n=8$, $G_2$ for $n=7$, $\SU(3)$ for $n=6$ and $\Sp(1) = \SU(2)$
for $n=4$.  In all cases we obtain that $[\fg_1,\fg_1]$ is the
translation ideal $\RR^{1,10-n}$.

\subsection{Branes}
\label{sec:branes}

For the elementary half-BPS M2- and M5-brane backgrounds one also
finds that $[\fg_1,\fg_1]$ is the translation ideal $\RR^{1,p}$ on the
brane.  Both backgrounds are geometrically a warped product
\begin{equation*}
  g = H^\alpha \eta + H^\beta \delta~,
\end{equation*}
where $\eta$ is the Minkowski metric on $\RR^{1,p}$, $p=2,5$;
$\delta$ is the Euclidean metric on $\RR^q$, $q=8,5$, respectively;
and $H$ is a harmonic function on $\RR^q$ such that the metric is
asymptotically flat.  The coefficients $\alpha$ and $\beta$ are given
in terms of $p$, but we do not need their explicit form.  The Killing
spinors are given by
\begin{equation*}
  \varepsilon = H^{\alpha/4} \varepsilon_\infty~,
\end{equation*}
where $\varepsilon_\infty$ is a parallel spinor in the asymptotically
flat geometry which obeys the algebraic condition
\begin{equation*}
  \nu_\eta \cdot \varepsilon_\infty = \varepsilon_\infty~,
\end{equation*}
where $\nu_\eta$ is the volume form of the Minkowski metric $\eta$.
Notice that the same identity is satisfied by $\varepsilon$ itself.

Consider the case of the M2-brane.  Here $\nu_\eta$ is a $3$-form
and hence it is self-adjoint relative to the symplectic structure on
the spinor bundle.  Let $X$ be perpendicular to the brane
world-volume.  Then $X \cdot \nu_\eta = - \nu_\eta \cdot X$, and hence
if $\varepsilon_1$ and $\varepsilon_2$ are Killing spinors,
\begin{align*}
  (\varepsilon_1, X \cdot \varepsilon_2) &= 
  (\varepsilon_1, X \cdot \nu_\eta \cdot \varepsilon_2)\\
  &= (\varepsilon_1, - \nu_\eta \cdot X \cdot \varepsilon_2)\\
  &= - (\nu_\eta \cdot \varepsilon_1, X \cdot \varepsilon_2)\\
  &= - (\varepsilon_1, X \cdot \varepsilon_2)~.
\end{align*}
Therefore $\xi(\varepsilon_1,\varepsilon_2)$ is in the double
perpendicular of the tangent space to the world-volume of the brane,
whence tangent to the world-volume of the brane.  This result is
intuitively obvious because this argument works for any harmonic
function $H$, even if this function has no symmetries.

A similar calculation shows the analogous result for the M5-brane.
Here $\nu_\eta$ is a $6$-form, whence it is symplectically
skew-adjoint.  However, if $X$ is perpendicular to the brane
world-volume, now $X \cdot \nu_\eta = \nu_\eta \cdot X$.  A calculation
virtually identical to the one above yields that
$\xi(\varepsilon_1,\varepsilon_2)$ is tangent to the brane
world-volume.

Now if $X$ is tangent to the brane world-volume and $\varepsilon$ is a
Killing spinor, a quick calculation shows that
\begin{equation*}
  \nabla_X \varepsilon = \tfrac14 \alpha d\log H \cdot X \cdot
  \varepsilon~.
\end{equation*}
Let $Y = Y_{||} + Y_\perp$ be any vector field, where we have
decomposed into parallel and perpendicular components with respect to
the brane world-volume, and let $\varepsilon_1,\varepsilon_2$ be
Killing spinors.  Then,
\begin{align*}
  \left< \nabla_X \xi(\varepsilon_1,\varepsilon_2), Y \right> &=
  \left< \nabla_X \xi(\varepsilon_1,\varepsilon_2), Y_{||} \right>\\
  &= \left< \xi(\nabla_X \varepsilon_1,\varepsilon_2), Y_{||} \right>
  + \left< \xi(\varepsilon_1,\nabla_X \varepsilon_2), Y_{||} \right>\\
  &= (\nabla_X \varepsilon_1, Y_{||} \cdot \varepsilon_2) +
  (\varepsilon_1, Y_{||} \cdot \nabla_X \varepsilon_2)\\
  &= \tfrac14 \alpha (d\log H \cdot X \cdot \varepsilon_1, Y_{||}  \cdot
  \varepsilon_2) + \tfrac14 \alpha (\varepsilon_1, Y_{||} \cdot d\log H
  \cdot X \cdot \varepsilon_2)\\
  &= \tfrac14 \alpha (d\log H \cdot \varepsilon_1, X \cdot Y_{||}
  \cdot \varepsilon_2 ) + \tfrac14 \alpha (d\log H \cdot
  \varepsilon_1, Y_{||} \cdot X \cdot \varepsilon_2 )\\
  &= \half \alpha \left<X,Y\right> (\varepsilon_1, d\log H \cdot
  \varepsilon_2)\\
  &= \half \alpha \left<X,Y\right> \left<
    \xi(\varepsilon_1,\varepsilon_2), d \log H\right>\\
  &= 0~,
\end{align*}
where we have used repeatedly that $d\log H$ is perpendicular to the
brane world volume.  In other words, the Lorentz component of
$\xi(\varepsilon_1,\varepsilon_2)$ vanishes, whence it is a
translation.

Virtually the same argument applies for the M2-brane at a conical
singularity \cite{AFHS}, where the transverse euclidean metric
$\delta$ is replaced by a cone of holonomy contained in $\Spin(7)$.

A similar argument also works if we curve the world-volume as in
\cite{BP} and \cite{JMRF}.  In the case of static brane world-volumes
\cite{BP}, $[\fg_1,\fg_1]$ contains the timelike parallel vector,
whereas in the case of the indecomposable supersymmetric waves
\cite{JMRF}, $[\fg_1,\fg_1]$ once again coincides with the null
parallel vector.

The Killing superalgebras of the Freund--Rubin backgrounds which
appear as near-horizon geometries of elementary branes and of branes
at conical singularities have been described in \cite{AFHS}.

\section{Supersymmetry and homogeneity}
\label{sec:susyhom}

In this section we will prove that 24+ backgrounds are locally
homogeneous.

\subsection{Homogeneous backgrounds}
\label{sec:hombgs}

We recall that a background $(M,g,F)$ is homogeneous if the group $G$
of $F$-preserving isometries acts transitively on $M$, so that for any
two points $p,q\in M$, there is a $g\in G$ such that $q = g \cdot p$.
In supergravity we usually work with local metrics and do not
necessarily impose completeness of the background.  In this context,
the relevant concept is not homogeneity but \emph{local} transitivity,
namely that every $p \in M$ is contained in a neighbourhood $U$ such
that for every $q \in U$ there is a local $F$-preserving isometry $g$
such that $q = g\cdot p$.  This is equivalent to the existence of a
frame consisting of $F$-preserving Killing vectors at every point
$p\in M$.  This implies that the background is locally homogeneous;
that is, that given any $p,q \in M$ there are neighbourhoods $U$ of
$p$ and $V$ of $q$ and a local $F$-preserving isometry $g: U \to V$
such that $g\cdot p = q$.
\begin{proof}
  Since $M$ is connected, let $\gamma: I=[0,1] \to M$ be a continuous
  curve connecting $p$ and $q$.  For every $t\in I$, there is a
  neighbourhood $U_t \subset M$ of $\gamma(t)$ with the property that
  every $r \in U_t$ is related to $\gamma(t)$ by a local
  $F$-preserving isometry.  The intersections $U_t \cap \gamma(I)$
  define an open cover (relative to the subspace topology) for
  $\gamma(I)$.  Since $\gamma$ is continuous, the preimages $V_t =
  \gamma^{-1}(U_t \cap \gamma(I))$ are an open cover of the interval.
  Since the interval is compact, there is a finite subcover $V_i =
  V_{t_i}$ for $i=0,\dots, N$ for some $N$.  We can further choose
  that $0=t_0 < t_1 < \dots < t_N = 1$ and that the successive
  intersections $V_i \cap V_{i+1}$ are nonempty.  Then choose $r_i \in
  \gamma(V_{i-1}) \cap \gamma(V_i)$.  By hypothesis, there exist local
  isometries $g_i,h_i$ such that $g_i \gamma(t_i)= r_i$ and $h_i
  \gamma(t_i) = r_{i+1}$.  The desired local isometry between $p$ and
  $q$ is given by
  \begin{equation*}
    \psi := g_N^{-1} \circ h_{N-1} \circ \dots \circ g_2^{-1} \circ
    h_1 \circ g_1^{-1} \circ h_0~.
  \end{equation*}
  Finally let $V$ be a small enough open neighbourhood of $q$, so that
  $U=\psi^{-1}(V)$ is defined.  The open set $U$ is a neighbourhood of
  $p$ and clearly $\psi: U \to V$.
\end{proof}

In the next section we will prove that any background admitting more
than 24 supersymmetries, so that $\nu > \frac34$, is locally
homogeneous.  We will prove this only using basic linear algebra, by
studying in detail the restriction of the map $\varphi_p$ defined in
\eqref{eq:varphi} to the subspace $W_p$ spanned by the values of the
Killing spinors at an arbitrary point $p\in M$ and showing that if
$\dim W_p > 24$ then the component $\xi_p$ of this map on $T_pM$ is
surjective.  We remark that this result is stronger than (local)
homogeneity in that we are proving that the symmetries which
\emph{follow} from the supersymmetry already act locally transitively.
For example, there are homogeneous plane waves admitting only 16
supersymmetries for which the only Killing vector which can be
constructed from Killing spinors is the parallel null vector; although
the Killing vectors for the homogeneous plane waves with more than 16
supersymmetries can be constructed from the Killing spinors
\cite{BMO}.  In fact, one can see that for all 16+ solutions mentioned
in the introduction local transitivity is already implied by
supersymmetry.

\subsection{Local homogeneity of 24+ backgrounds}
\label{sec:24+hom}

We will fix a point $p\in M$ once and for all.  The tangent space
$T_pM$ with the restriction of the metric $g(p)$ becomes a lorentzian
inner product space.  We will denote it $V$ and will let
$\left<-,-\right>$ denote the lorentzian inner product and $|-|^2$
denote the associated (indefinite) norm.  The fibre $\eS_p$ of the
spinor bundle is isomorphic to the irreducible $\Cl(V)$-module $S$,
which is a 32-dimensional real symplectic vector space, with
symplectic structure denoted by $(-,-)$ as above.  Let $W \subset S$
be the subspace corresponding to the Killing spinors.  The map
\eqref{eq:square} defines a symmetric bilinear map
\begin{equation*}
  \xi: S \otimes S \to V~.
\end{equation*}
We want to show that if $\dim W$ is large enough, then the restriction
\begin{equation*}
  \xi|_W : W \otimes W \to V
\end{equation*}
of $\xi$ to $W$ is surjective.  This means that $T_pM$ is spanned
by the values of $F$-preserving Killing vectors.  Since $p$ is
arbitrary, this will be the case at every point and the background
will be locally homogeneous.

Clearly if $W = S$ then $\xi$ is surjective: this follows from the
representation theory of the spin group.  On the other hand there are
examples with $\dim W = 16$ which are not homogeneous, hence there has
to be a minimal $16 < N \leq 32$ such that whenever $\dim W \geq N$,
the map $\xi|_W$ is surjective.  We will show that $N=25$.

In our proof we will exploit the fact that $S$ is a symplectic vector
space, so it might be convenient to introduce some relevant notation
from symplectic linear algebra.  Let $W \subset S$ be any vector
subspace.  The vectors which are symplectically perpendicular to all
the vectors in $W$ define a subspace
\begin{equation*}
  W^\perp = \left\{ \varepsilon \in S \mid (\varepsilon,w) = 0 \quad
    \text{for all $w\in W$} \right\}~.
\end{equation*}
Analogous to the case of a euclidean structure, we also have that
\begin{equation}
  \label{eq:complementarity}
  \dim W + \dim W^\perp = \dim S~,
\end{equation}
even though $W$ and $W^\perp$ are not generally disjoint.  For
example, every one-dimensional subspace is contained in its symplectic
perpendicular.  The relationship between $W$ and $W^\perp$ defines
certain types of subspaces.  For example, a subspace such that $W
\subset W^\perp$ is called \emph{isotropic}.  Clearly the dimension of
an isotropic subspace is at most half the dimension of $S$.  When the
dimension is precisely half, so that $W = W^\perp$, $W$ is called
\emph{lagrangian}.  At the other extreme, if $W$ and $W^\perp$ are
disjoint, then $W$ is said to be a \emph{symplectic} subspace.

As a side remark, we mention the intriguing fact that the Killing
spinors for the elementary half-BPS backgrounds define special
subspaces: lagrangian in the case of the M5-brane and the M-wave and
symplectic in the case of the M2-brane and the Kaluza--Klein monopole.
This is somewhat puzzling because the connection $\eD$ does not
preserve the symplectic structure in general.  It does for the purely
gravitational backgrounds, whose holonomy group is contained in
$\Spin(1,10)$ and hence in $\Sp(32,\RR)$, but the calculations in
\cite{BDLWHol} show that the holonomy algebras of the M2-brane and
M5-brane are not contained in the symplectic subalgebra
$\fsp(32,\RR)$.\footnote{We are grateful to George Moutsopoulos for
  checking this.}

Let us now proceed with the proof.  Let $\dim W > 16$, since the known
examples already negate anything else.  The map $\xi|_W$ is surjective
if and only if the subspace perpendicular to its image is trivial.
Equivalently, if and only if the only vector $v \in V$ obeying
\begin{equation}
  \label{eq:condition}
  (\varepsilon_1, v \cdot \varepsilon_2) = 0\qquad \text{for all
    $\varepsilon_i \in W$}
\end{equation}
is the zero vector $v=0$.  Throughout this section we will allow
vectors (and not just forms) to act on spinors.  By definition, the
action of a vector $v$ is simply the Clifford action of the dual
one-form $v^\flat$.

Our first observation is that any $v\in V$ satisfying
\eqref{eq:condition} is necessarily null.  Indeed, notice that
\eqref{eq:condition} can be rephrased as saying that as a Clifford
endomorphism
\begin{equation*}
  v : W \to W^\perp~.
\end{equation*}
Since $\dim W > \half \dim S$, it follows from
\eqref{eq:complementarity} that $\dim W > \dim W^\perp$, whence $v$
must have kernel, purely on dimensional grounds.  On the other hand,
the Clifford algebra says that $v^2 = -|v|^2 \1$, whence $v$ has
kernel if and only if $|v|^2 = 0$.

Since in a lorentzian vector space all null subspaces are
one-dimensional, we deduce that the subspace perpendicular (relative
to $\left<-,-\right>$) to the image of $\xi$ is at most
one-dimensional.  Moreover, if one-dimensional, it is spanned by a
null vector $v \in V$.

Our next step is to show that in this case the Clifford endomorphism
$v$ has rank $16$.  From $v^2 = 0$, we see that $\im v \subset \ker
v$.  To show the reverse inclusion, let $u \in V$ be a complementary
null vector such that
\begin{equation*}
  u \cdot v + v \cdot u =  \1~.
\end{equation*}
(In other words, we can think of $v$ as $\Gamma_+$ and $u$ as
$\Gamma_-$.)  Then applying both sides of this identity to a vector
$\varepsilon$ annihilated by $v$, we find
\begin{equation*}
  \varepsilon = v \cdot u \cdot \varepsilon  \in \im v~,
\end{equation*}
whence $\ker v = \im v$.  A similar argument shows that $\ker u = \im
u$.  Moreover, from \eqref{eq:cliffskew} it follows that $\ker u$ and
$\ker v$ are complementary lagrangian subspaces of $S$.  In
particular, $\rank v = \dim\im v = 16$.

Now let $U$ be a complementary subspace to $W$, so that $S = W \oplus
U$.  Relative to this split, the symmetric bilinear form $\beta$,
defined by
\begin{equation*}
  \beta(\varepsilon_1, \varepsilon_2) = (\varepsilon_1, v \cdot
  \varepsilon_2)~,
\end{equation*}
has the following matrix
\begin{equation*}
  \begin{pmatrix}
    0 & A\\ A^t & B
  \end{pmatrix}~,
\end{equation*}
where $A: U \to W$, $A^t: W \to U$ and $B: U \to U$ are linear maps.
We know that this matrix has rank $16$, since $(-,-)$ is nondegenerate
and $v$ has rank $16$.  What we will do now is estimate the maximal
possible rank in terms of the dimension of $W$.

The kernel of $\beta$ consists of $(w,u)\in W \oplus U$ such that
$Au=0$ and $A^t w + Bu = 0$.  Notice that $\dim U < \dim W$, whence
$\rank A \leq \dim U$.  In the case of maximal rank, the only solution
of $Au=0$ is $u=0$.  In this case the kernel of $\beta$ consists of
$(w,0)$ with $w\in\ker A^t$.  In other words, the dimensions of the
kernels of $\beta$ and of $A^t$ agree.  Since $A^t$ and $A$ have the
same rank, $A^t$ is onto, whence its kernel has dimension $\dim W -
\dim U$.  Therefore the rank of $\beta$ is \emph{at most} $32 - \dim W
+ \dim U = 2 \dim U$; but we know that the rank of $\beta$ is 16,
whence $16 \leq 2 \dim U$ or $\dim U \geq 8$.  This means that if
$\dim U < 8$ (equivalently, if $\dim W > 24$) no such $v$ can exist
and the map $\xi|_W$ is surjective.

\section{Conclusions}
\label{sec:conc}

In this paper we have investigated the relation between symmetry and
supersymmetry in supergravity backgrounds, concentrating for
definiteness in eleven-dimensional supergravity.  We have shown that
the Killing spinors in any such background generate a Lie
superalgebra.  Strictly speaking they generate an ideal of what we
call the Killing superalgebra of the background, which may contain
additional ``accidental'' bosonic symmetries.  The Killing
superalgebra has appeared before in many special cases, but until now
there was no general proof that this construction resulted in a Lie
superalgebra.

Since supersymmetries generate symmetries, we posed the general
question of whether there is a mininum amount of supersymmetry that a
solution must preserve for it to be automatically (locally)
homogeneous.  Homogeneous backgrounds are particularly tractable and a
positive answer to that question implies that a classification of
homogeneous backgrounds, for example, would automatically imply a
classification of solutions preserving more than a certain critical
fraction $\nu_c$ of supersymmetry.

We have reviewed what is known about backgrounds admitting more than
16 supersymmetries and have observed that all known such backgrounds
are homogeneous.  Moreover, it is the ideal of the Killing
superalgebra generated by the supersymmetries which already acts
locally transitively.  We have checked this for the known 16+
solutions and also for some recently discovered ones, included in the
appendices.

Finally we have proven that if a solution preserves more than 24
supersymmetries then the ideal of the Killing superalgebra generated
by these supersymmetries acts locally transitively on the background.
In particular, these 24+ backgrounds are locally homogeneous.

\section*{Acknowledgments}

The research of JMF is partially funded by the EPSRC grant
GR/R62694/01.  Part of this work was written during a visit of JMF to
the Erwin Schrödinger Institute to participate in the programme on
\emph{String theory on curved backgrounds} and a subsequent visit to
the IHÉS to participate in the \emph{Avant Strings} workshop.  It is a
pleasure to thank both institutions for their support and hospitality,
and especially Andreas Recknagel and Jean-Pierre Bourguignon for the
respective invitations.

PM would like to thank Luis Alvarez-Gaumé, Matthias Blau, Bert
Janssen, Martin O'Loughlin and Tomás Ortín for fruitful discussions
and Josefina Millán Jiménez for her (library) support.  PM would like
to thank el Instituto de Física Teórica (Madrid) and het Instituut
voor Theoretische Fysica (Leuven) for their support and hospitality.

The research of SP is funded in part by an EPSRC Postgraduate
studentship.

\appendix

\section{Clifford algebra conventions}
\label{app:clifford}

Our Clifford algebra conventions mostly follow the book \cite{LM}, but
we will review them here briefly.

Let $\RR^{s,t}$ denote the real $(s+t)$-dimensional vector space with
inner product obtained from the norm
\begin{equation*}
  |\bx|^2 = -(x^1)^2 - \cdots - (x^t)^2 + (x^{t+1})^2 + \cdots +
  (x^{t+s})^2~,
\end{equation*}
for $\bx=(x^1,\dots,x^{s+t}) \in \RR^{s,t}$.  By definition the real
Clifford algebra $\Cl(s,t)$ is generated by $\RR^{s,t}$ (and the
identity $\1$) subject to the Clifford relation
\begin{equation*}
  \bx \cdot \bx  = - |\bx|^2 \1~,
\end{equation*}
where we ask the reader to pay close attention to the sign!

We are interested in eleven-dimensional lorentzian signature:
$\RR^{1,10}$.  As a real associative algebra, $\Cl(1,10)$ is
isomorphic to two copies of the algebra of $32 \times 32$ real
matrices.  This means that there are (up to isomorphism) two
irreducible representations $\fM_\pm$, which are real and thirty-two
dimensional.  They are distinguished by the action of the generator of
the centre of $\Cl(1,10)$, which is realised geometrically by the
volume form $\boldsymbol{\nu}$ of $\RR^{1,10}$.

The Clifford algebra $\Cl(1,10)$ is isomorphic as a real vector space
(but \emph{not} as an algebra) to the exterior algebra
$\Lambda\RR^{1,10}$.  In this way, elements of $\Lambda\RR^{1,10}$ can
act on $\fM_\pm$.

Now let $(M,g)$ be a lorentzian eleven-dimensional manifold, with
signature $(1,10)$.  We can choose local orthonormal frames for the
tangent bundle $TM$ and dual coframes for the cotangent bundle $T^*M$.
Relative to such a coframe, each cotangent space is isomorphic to
$\RR^{1,10}$ as an inner product space and we can construct at each
point a Clifford algebra $\Cl(1,10)$.  As we let the point vary, these
algebras patch up nicely to yield a bundle $\Cl(T^*M)$ of Clifford
algebras which, as a vector bundle, is isomorphic to $\Lambda T^*M$.
The isomorphism $\Lambda\RR^{1,10} \cong \Cl(1,10)$ also extends to
give a bundle isomorphism $\Lambda T^*M \cong \Cl(T^*M)$.

If in addition $(M,g)$ is spin, then there are (not necessarily
unique) vector bundles $\eS_\pm$ associated to each the irreducible
representations $\fM_\pm$ of $\Cl(1,10)$.  These are bundles of
modules over the Clifford bundle $\Cl(T^*M)$.  Differential
forms---that is, sections of $\Lambda T^*M$---act naturally on
sections of $\eS_\pm$ via the isomorphism $\Lambda T^*M \to \Cl(T^*M)$
and the natural pointwise action of $\Cl(T^*M)$ on $\eS_\pm$.

In this paper we will have ample opportunity to compute Clifford
products of differential forms acting on sections of $\eS_\pm$.  We
collect here some useful formulae.

If $X$ is a vector and $\omega$ a $p$-form, then
\begin{equation}
  \label{eq:Xomega}
  X^\flat \cdot \omega = X^\flat \wedge \omega - \iota_X \omega~,
\end{equation}
and
\begin{equation}
    \label{eq:omegaX}
  \omega \cdot X^\flat = (-1)^p \left( X^\flat \wedge \omega + \iota_X
    \omega\right)~,
\end{equation}
where $\flat$ is the musical isomorphism from vectors to one-forms
induced by the metric; that is, the one-form $X^\flat$ is defined by
$X^\flat(Y) = \left<X,Y\right>$ for every vector $Y$.  Iterating these
identities we find, for example,
\begin{equation}
  \label{eq:omegaXY}
  \omega \cdot (X^\flat \wedge Y^\flat) = X^\flat \wedge Y^\flat
  \wedge \omega + \iota_X \iota_Y \omega + X^\flat \wedge \iota_Y
  \omega -  Y^\flat \wedge \iota_X \omega~,
\end{equation}
and
\begin{equation}
  \label{eq:XYomega}
  (X^\flat \wedge Y^\flat) \cdot \omega = X^\flat \wedge Y^\flat
  \wedge \omega + \iota_X \iota_Y \omega - X^\flat \wedge \iota_Y
  \omega + Y^\flat \wedge \iota_X \omega~.
\end{equation}
If $\omega$ is a $p$-form and $\star \omega$ its Hodge dual, then
their Clifford actions are related by
\begin{equation}
  \label{eq:hodge}
  \star \omega = (-1)^{p(p+1)/2} \omega \cdot \boldsymbol{\nu}~,
\end{equation}
where $\boldsymbol{\nu}$ is the volume form.

The bundles $\eS_\pm$ inherit from $\fM_\pm$ a symplectic structure
which is compatible with the action of the Clifford algebra; that is,
the Clifford endomorphisms corresponding to 1-forms (equivalently,
vectors) are skew-symmetric:
\begin{equation}
  \label{eq:cliffskew}
  (\varepsilon_1, v^\flat \cdot \varepsilon_2) = -   (v^\flat \cdot
  \varepsilon_1, \varepsilon_2)~.
\end{equation}
In turn, this identity implies that the bilinear form
\begin{equation*}
  \beta_v (\varepsilon_1, \varepsilon_2) = (\varepsilon_1,
  v^\flat \cdot \varepsilon_2)
\end{equation*}
associated to the vector $v$ is symmetric.

More generally, if $\omega$ is a $p$-form, we will let $\omega^*$
denote its adjoint with respect to this symplectic structure; that is,
\begin{equation}
  \label{eq:sympadj}
  (\omega \cdot \varepsilon_1, \varepsilon_2) = 
  (\varepsilon_1, \omega^* \cdot \varepsilon_2)~.
\end{equation}
Explicitly, one finds that
\begin{equation}
  \label{eq:adjoint}
  \omega^* = (-1)^{p(p+1)/2} \omega~,
\end{equation}
whence 1-forms, 2-forms and 5-forms (and their Hodge duals) preserve
the symplectic structure.  Indeed, $\fsp(32,\RR) = \Lambda^1 \oplus
\Lambda^2 \oplus \Lambda^5$ under $\fso(1,10)$.

\section{Homogeneity of some 16+ discrete quotients}
\label{app:homogquots}

The possible Kaluza--Klein reductions (by one-parameter subgroups) of
the maximally supersymmetric Freund--Rubin backgrounds of
eleven-dimensional and type IIB supergravities have been classified in
\cite{FigSimAdS}.  Associated to these reductions, there are discrete
quotients by a cyclic subgroup.  Two of these reductions gave rise to
backgrounds with more than 16 supercharges and the same is true for
the associated quotients.  In this appendix we will show that for
every $N > 1$ there is a $\ZZ_N$-quotient $\AdS_4 \times (S^7/\ZZ_N)$
with 24 supercharges, and that there is a a two-parameter family of
$\ZZ$-quotients $(\AdS_4 \times S^7)/\ZZ$ with 18 supercharges.  We
will then demonstrate that these quotients remain homogeneous.

\subsection{A family of $\nu=\frac34$ quotients}
\label{appsec:fam34}

This family of backgrounds has the form $\AdS_4 \times (S^7/\Gamma)$
where $\Gamma$ is a finite cyclic subgroup of $\SO(8)$.  Since
$\SO(8)$ is compact, the exponential map is surjective and $\Gamma$
will be generated by an element $\gamma$ in the image of the
exponential map.  Let us identify the Lie algebra $\fso(8)$ with the
$8 \times 8$ skew-symmetric real matrices.  Then consider the element
$J \in \fso(8)$ given by
\begin{equation}
  \label{eq:J}
  J = 
  \begin{pmatrix}
    0 & 1 &  &  &  &  &  & \\
    -1 & 0 &  &  &  &  &  & \\
     &  & 0 & 1 &  &  &  & \\
     &  & -1 & 0 &  &  &  & \\
     &  &  &  & 0 & 1 &  & \\
     &  &  &  & -1 & 0 &  & \\
     &  &  &  &  &  & 0 & -1\\
     &  &  &  &  &  & 1 & 0
  \end{pmatrix}~.
\end{equation}
Then consider $\gamma = \exp(2\pi J/N) \in \SO(8)$.  Explicitly,
\begin{equation*}
  \gamma = 
  \begin{pmatrix}
    \cos\frac{2\pi}N & \sin\frac{2\pi}N &  &  &  &  &  & \\
    -\sin\frac{2\pi}N & \cos\frac{2\pi}N &  &  &  &  &  & \\
     &  & \cos\frac{2\pi}N & \sin\frac{2\pi}N &  &  &  & \\
     &  & -\sin\frac{2\pi}N & \cos\frac{2\pi}N &  &  &  & \\
     &  &  &  & \cos\frac{2\pi}N & \sin\frac{2\pi}N &  & \\
     &  &  &  & -\sin\frac{2\pi}N & \cos\frac{2\pi}N &  & \\
     &  &  &  &  &  & \cos\frac{2\pi}N & -\sin\frac{2\pi}N\\
     &  &  &  &  &  & \sin\frac{2\pi}N & \cos\frac{2\pi}N
  \end{pmatrix}~.
\end{equation*}
Clearly $\gamma^N = 1$ and hence it generates a $\ZZ_N$ subgroup of
$\SO(8)$.  This subgroup acts freely on $S^7$ and the resulting
quotient is a smooth lens space.  The element
\begin{equation*}
  \widehat\gamma = \exp(\tfrac{\pi}N \Gamma_{12}) \exp(\tfrac{\pi}N
  \Gamma_{34}) \exp(\tfrac{\pi}N \Gamma_{56}) \exp(-\tfrac{\pi}N
  \Gamma_{78}) \in \Spin(8)
\end{equation*}
is the spin lift of $\gamma$ and clearly obeys $\widehat\gamma^N = 1$,
whence as explained in \cite{FigSimAdS,FOMRS}, $\Gamma$ lifts to
$\Spin(8)$ making the quotient lens space into a spin manifold.

The Killing spinors which survive to the quotient are the
$\Gamma$-invariant Killing spinors on $\AdS_4 \times S^7$.  As
reviewed in \cite{FigLeiSim} for general Freund--Rubin backgrounds,
the Killing spinors on $\AdS_4 \times S^7$ are given by tensor
products of geometric Killing spinors of $\AdS_4$ and $S^7$.  Since
$\Gamma$ only acts on the sphere, we will concentrate on the sphere.
The cone construction of \cite{Baer} relates the geometric Killing
spinors on $S^7$ to a chiral spinor representation of $\Spin(8)$.  In
our conventions, this is the spinor representation of negative
chirality, which under the action of $\gamma$ is seen to have six zero
weights, as explained in more detail in
\cite[Section~6.2.1]{FigSimAdS}.  This implies that the quotient
background preserves a fraction $\nu = \frac68 = \frac34$ of the
supersymmetry.

We now show that the quotient is homogeneous, but first some general
remarks.  Let $(M,g)$ have isometry group $G$ and let $\Gamma \subset
G$ be a discrete subgroup acting freely on $M$ with a smooth quotient
$M/\Gamma$.  Not all the isometries of $M$ will descend to isometries
in the quotient.  Indeed, a necessary and sufficient condition for an
isometry $g\in G$ to act on $M/\Gamma$ is that if two points $p,q \in
M$ are in the same $\Gamma$-orbit, so are their images $g \cdot p, g
\cdot q$.  The subgroup of $G$ thus defined is the normaliser
\begin{equation*}
  N_\Gamma = \left\{ g \in G \mid g \gamma g^{-1} \in
      \Gamma\quad\forall \gamma \in \Gamma \right\}
\end{equation*}
of $\Gamma$ in $G$.  Since $\Gamma$ is discrete, the connected
component of $N_\Gamma$ containing the identity is the centraliser
\begin{equation*}
  Z_\Gamma = \left\{ g \in G \mid g \gamma g^{-1} = \gamma\quad\forall
    \gamma \in \Gamma \right\}
\end{equation*}
of $\Gamma$ in $G$.  To see this, simply consider a path $g(t)$ from
the identity to $g$ in the same connected component of $N_\Gamma$ and
consider its action on any $\gamma \in \Gamma$.  Since $g(t) \in
N_\Gamma$ for all $t$, we have that $g(t) \gamma g(t)^{-1} \in \Gamma$
for all $t$.  Since $\Gamma$ is discrete, continuity means that this
has to be the same element of $\Gamma$ for all $t$, but it is $\gamma$
itself when $t=0$.

For the case at hand, the centraliser $Z_\Gamma$ is the subgroup of
$\SO(8)$ which commutes with the matrix $J$ in \eqref{eq:J}.  Now $J$
is a complex structure and the subgroup thus defined is isomorphic to
$\U(4)$, which still acts transitively on $S^7$ (with isotropy
$\U(3)$) and hence will continue to do so in the lens space
$S^7/\Gamma$.  In other words, $S^7/\Gamma$ is homogeneous, and hence
so is $\AdS_4 \times (S^7/\Gamma)$. 

The Killing superalgebra of the above solution must be a
sub-superalgebra of the superalgebra $\fosp(8|2,\RR)$
corresponding to $\AdS_4\times S^7$.  In fact, it is not hard to
see that the superalgebra is $\fu(1)\oplus\fosp(6|2,\RR)$,
which is a regular maximal sub-superalgebra of $\fosp(8|2,\RR)$
\cite{DictLieSuperAlg}.  This means that only the $\fsu(4)$ is
generated by Killing spinors, but since this acts transitively on
$S^7$, and will continue to do so on $S^7/\Gamma$, we see that
also in this case supersymmetry is responsible for homogeneity.

\subsection{A family of $\nu = \frac9{16}$ quotients}
\label{appsec:fam916}

This family of quotients is slightly more involved than the previous
one, since the group $\Gamma$ defining the quotient acts on both
$\AdS_4$ and on $S^7$.  Let $J \in \fso(8)$ be as in \eqref{eq:J}.
The isometry algebra of $\AdS_4$ is $\fso(2,3)$, which we can identify
with the $5\times 5$ real matrices which are skew-symmetric relative
to a metric $\eta$ of signature $(2,3)$.  Let us take $\eta$ to be
diagonal with entries $(-1,-1,1,1,1)$ and let $L \in \fso(2,3)$ be the
following matrix
\begin{equation}
  \label{eq:L}
  L = 
  \begin{pmatrix}
    0 & 1 & 1 & 0 & 0\\
    -1 & 0 & 0 & -1 & 0\\
    1 & 0 & 0 & 1 & 0\\
    0 & -1 & -1 & 0 & 0\\
    0 & 0 & 0 & 0 & 0
  \end{pmatrix}~.
\end{equation}
Now let $\alpha,\beta$ be positive real numbers and consider the
element
\begin{equation*}
  \gamma = \exp(\alpha L + \beta J) \in \widetilde{\SO}(2,3) \times
  \SO(8)~,
\end{equation*}
where $\widetilde{\SO}(2,3)$, the isometry group of $\AdS_4$, is an
infinite cyclic cover of $\SO(2,3)$, as discussed in detail in
\cite[Section~5.1.2]{FigSimAdS}.  This element $\gamma$ generates an
infinite cyclic subgroup which, as shown in
\cite[Section~6.2.4]{FigSimAdS}, acts freely on $\AdS_4 \times S^7$.
The calculations in \cite[Section~6.2.4]{FigSimAdS}, though written
for the continuous $\RR$-quotients, show that these $\ZZ$-quotients
preserve a fraction $\nu = \frac9{16}$ of the supersymmetry.  Indeed,
notice that $J$ is as above and we have already seen that it preserves
$\frac34$ of the $S^7$-supersymmetry.  Notice that $L^2=0$ and that
the same analysis as in \cite[Section~6.2.4]{FigSimAdS} shows that $L$
preserves another $\frac34$ of the $\AdS_4$ supersymmetry.  Thus we
see that there are $6$ $\Gamma$-invariant Killing spinors on $S^7$ and
$3$ on $\AdS_4$ for a total of $18$ supercharges in the quotient.

It is however clear that there are, as far as supersymmetry is 
concerned, two special cases in this family: $\alpha =0$ and $\beta =0$. 
In both cases the fraction of preserved supersymmetry is 
$\nu =\textstyle{\frac{3}{4}}$, and the geometry corresponds to 
$\AdS_4\times S^7/\Gamma$, respectively $\AdS_4/\Gamma\times S^7$.
This last case was not treated in \cite{FigSimAdS} since the associated
Killing vector has zero norm, whereas \cite{FigSimAdS} focuses on spacelike
quotients.

To show homogeneity of the quotient we proceed as before and show that
the centraliser $Z_\Gamma$ acts transitively already before taking the
quotient.  The centraliser is the product of the centralisers of the
projections of $\Gamma$ to $\widetilde{\SO}(2,3)$ and $\SO(8)$
respectively.  We already know from the previous section that the
$\SO(8)$-factor is $\U(4)$.  The $\widetilde{\SO}(2,3)$-factor of the
centraliser is easier to describe infinitesimally; that is, we will
describe its Lie algebra which has the form of a semidirect product
\begin{equation*}
  \fk = \fsl(2,\RR)_+ \ltimes \fh_3~,
\end{equation*}
where $\fsl(2,\RR)_+$ is the self-dual $\fsl(2,\RR)$ in $\fso(2,2)
\cong \fsl(2,\RR)_+ \oplus \fsl(2,\RR)_-$ and $\fh_3$ is a
three-dimensional Heisenberg algebra where the central element is
precisely the element $L$ in \eqref{eq:L}.  (Notice that $L$ belongs
to $\fsl(2,\RR)_-$.)  It is not hard to show that the subgroup
$K\subset \SO(2,3)$ with Lie algebra $\fk$ acts transitively on the
quadric $-x_1^2 - x_2^2 + x_3^2 + x_4^2 + x_5^2 = - R^2$ in
$\RR^{2,3}$, whence its infinite cyclic cover $\widetilde K\subset
\widetilde{\SO}(2,3)$, obtained by extending $K$ by the fundamental
group of the quadric, also acts transitively on $\AdS_4$.  In summary,
$\widetilde{K} \times \U(4)$ acts transitively on $\AdS_4 \times S^7$
and hence does so on the quotient.

The Killing superalgebra can readily be found by projection, but for
definiteness let us discuss the case $\beta =0$.  It can then be seen
that under $\fsl(2,\RR)\oplus \fso(8) \subset \fso(2,3)\oplus
\fso(8)$, the invariant Killing spinors transform as
$(\repre2,\repre8)\oplus (\repre1,\repre8)$.  The
$\fsl(2,\RR)$-singlets combine with the $\fh_{3}$ into a Heisenberg
superalgebra $\fh_{3|8}$, meaning that there are 2 bosonic and 8
fermionic creation and annihilation generators, which is a super-ideal
of the full superalgebra.  The $(\repre2,\repre8)$-spinors together
with the $\fso(8)$ and the $\fsl(2,\RR)$ subalgebras, can be seen to
form the algebra $\fosp(8|1,\RR)$, making the full superalgebra
$\fosp(8|1,\RR) \ltimes \fh_{3|8}$.  It is clear that in the general
case, the necessary projections on the $\AdS_{4}$ and the $S^{7}$ part
are done independently, so that we can combine the above results with
the results in Appendix~\ref{appsec:fam34}, only to find that in the
generic case the Killing superalgebra is $\fu(1)\oplus\left(
  \fosp(6|1,\RR) \ltimes \fh_{3|6}\right)$.

\section{A plane wave solution with 22 supercharges}
\label{app:newwave}

In \cite{BMO} it was shown that the Penrose limit of the M-theory
Gödel solution generates a one parameter family of wave solutions that
interpolates between two Cahen--Wallach (CW) spaces. This family
generically preserves 20 supersymmetries which at one CW-point is
enhanced to 24.  The reasoning of \cite{BMO} can of course also be
applied to the other 16+ Gödel solutions presented in
\cite{HarTakaGoedel}, and for completeness we will discuss the
resulting non-symmetric plane wave solutions.

Reference \cite{HarTakaGoedel} finds three 16+ Gödel solutions: the
above mentioned M-theory Gödel solution which preserves 20
supersymmetries, an $n=4$ case which also preserves 20
supersymmetries, and finally the $n=5$ case which preserves 18
supersymmetries.  The Penrose limit of the $n=5$ is actually a
CW-space, and as such ought to be known.
 
The $n=4$ Gödel solution can be obtained form the type IIB maximally
supersymmetric BFHP wave \cite{NewIIB} by T-duality and oxidation to
eleven dimensions.  Its Penrose limit reads
\begin{gather*}
  g = 2du\left(dv - \left[ \half \beta^2 \vec{x}^2_{(3\rightarrow 8)}
      +2\beta^2 x_2^2 \right] du - 2\beta p\ x_2dx_1 \right)
  + d\vec{x}_{(1\rightarrow 9)}^2\\
  F = 2 \beta du \wedge \left( dx^{129} + p dx^{349} + p dx^{569} + p
    dx^{789} + 2 \sqrt{1-p^2} dx^{278} \right)~.
\end{gather*}

As for the solution in \cite{BMO} there are two values for which the
above solution becomes a Cahen--Wallach space: $p=0$ where the
solution preserves 24 supersymmetries, and $p=1$ where one finds 22
supersymmetries.  For $p>0$, the necessary projection operator onto
the extra supersymmetries is given by
\begin{equation*}
  \tfrac18 \left(1 + \sqrt{1-p^2} \gamma^{1789} - p \gamma^{9} \right)
  \left(3 + \gamma^{3456} + \gamma^{3478} + \gamma^{5678} \right)~,
\end{equation*}
where $\gamma$ are the generators of the transverse $\Cl(0,9)$ and
from which one can see that this solution has 6 extra supersymmetries,
for a total of 22.  For $p=0$ the relevant projector is $\half (1 +
\gamma^{1789})$, which shows that there are now 8 extra
supersymmetries.

\bibliographystyle{utphys}
\bibliography{AdS,AdS3,ESYM,Sugra,Geometry,CaliGeo,Algebra}

\end{document}